\documentclass[12pt,tightenlines,eqsecnum,floats,aps,amsmath,amssymb,nofootinbib,superscriptaddress,showpacs]{revtex4}

\usepackage{amsmath,amsthm,amssymb,amsfonts}

\usepackage{colordvi}
\newtheorem{thr}{Theorem}
\newtheorem{prop}[thr]{Proposition}

\newtheorem{df}{Definition}

\numberwithin{equation}{section}
\numberwithin{thr}{section}
\numberwithin{chr}{section}
\numberwithin{df}{section}

\begin{document}

\title{Flux-area operator and black hole entropy}

\author{J. Fernando \surname{Barbero G.}}
\email[]{fbarbero@iem.cfmac.csic.es} \affiliation{Instituto de
Estructura de la Materia, CSIC, Serrano 123, 28006 Madrid, Spain.}

\author{Jerzy  \surname{Lewandowski}}
\email[]{lewand@fuw.edu.pl}\affiliation{Instytut Fizyki
Teoretycznej, Uniwersytet Warszawski,  ul. Ho\.{z}a 69, 00-681
Warsaw, Poland.}

\author{Eduardo J. \surname{S. Villase\~nor}}
\email[]{ejsanche@math.uc3m.es} \affiliation{Instituto Gregorio Mill\'an, Grupo de Modelizaci\'on
y Simulaci\'on Num\'erica, Universidad Carlos III de Madrid, Avda.
de la Universidad 30, 28911 Legan\'es, Spain.} \affiliation{Instituto
de Estructura de la Materia, CSIC, Serrano 123, 28006 Madrid, Spain.}

\date{May 21, 2009}

\begin{abstract}
We show that, for space-times with inner boundaries, there exists a natural area operator different from the standard one used in loop quantum gravity. This new \textit{flux-area operator} has equidistant eigenvalues. We discuss the consequences of substituting the standard area operator in the Ashtekar-Baez-Corichi-Krasnov definition of black hole entropy by the new one. Our choice simplifies the definition of the entropy and allows us to consider only those areas that coincide with the one defined by the value of the level of the Chern-Simons theory describing the horizon degrees of freedom. We give a prescription to count the number of relevant horizon states by using spin components and obtain exact expressions for the black hole entropy. Finally we derive its asymptotic behavior, discuss several issues related to the compatibility of our results with the Bekenstein-Hawking area law and the relation with Schwarzschild quasi-normal modes.
\end{abstract}


\maketitle

\section{Introduction}

The study of black holes is a central topic in quantum gravity. In fact, the identification of the microscopic degrees of freedom accounting for the black hole entropy and the successful derivation of the Bekenstein-Hawking area law are important challenges that prospective quantum gravity theories must meet. Within the loop quantum gravity (LQG) formalism the identification of these degrees of freedom has been proposed in the classical papers of Ashtekar, Baez, Corichi and Krasnov \cite{ABCK,ABK}. The resulting framework will be referred to in the following as the ABCK approach.  The combinatorial problem of counting the relevant black hole states was recast in a rather manageable form by Domagala and Lewandowski in \cite{DL} and an effective way to use it to derive the Bekenstein-Hawking area law was found by Meissner \cite{M} (see also \cite{EF1}).

The Hilbert space of quantum states in the ABCK formalism is a tensor product of a volume Hilbert space corresponding to the quantum geometry in the bulk and a surface Hilbert space associated to the $U(1)$ Chern-Simons (CS) theory on the isolated horizon used to model a black hole in LQG. A somewhat disturbing feature of the initial ABCK proposal is the fact that the prequantized value of the horizon area (related to the level $\kappa$ of the CS theory according to $a_\kappa=4\pi\gamma\ell_P^2 \kappa$) does not belong to the spectrum of the standard area operator. This compatibility problem was taken care of in the original papers \cite{ABCK,ABK} by introducing a small area interval in the definition of the entropy, a procedure that resembles the standard way of introducing the microcanonical ensemble in Statistical Mechanics. An alternative to this is to pick a different choice for the area operator on the horizon. The basic idea is now to use some extra structure --that exists when an inner space-time boundary is introduced to model a black hole-- to define a different area operator that we will call $\hat{a}^{\rm flux}$. This new area operator has an equally-spaced spectrum containing the prequantized horizon area eigenvalues. It is important to highlight the fact that this area operator is an alternative choice \textit{within} LQG that is available when an inner boundary is introduced. It is also important to note that the matching between the CS theory on the horizon and gravity in the bulk is arguably more natural with the new choice of area operator and the study of the black hole entropy is notably simplified.

Our purpose here is to thoroughly discuss black hole entropy in this new framework. Our starting point will be the ABCK entropy definition. We will first justify how the change in the area operator allows us to count the microstates responsible for the entropy for a given value of the prequantized area without introducing an interval at this stage. After that we will find a way to recast the counting of physical states as a counting of third spin components in the spirit of \cite{DL}. This will simplify the computation of the entropy and shed some light about the issue of how this can be done in the general case\footnote{There are some issues related to the fact that the Domagala-Lewandowski prescription \cite{DL} in the case of the standard definition for the entropy does no use an interval but rather a sum involving all the area eigenvalues up to the considered black hole area.}. The details of the counting will be explored afterwards, in fact, we will be able to get exact, closed expressions for the entropy --involving hypergeometric functions-- that can be effectively used to obtain the asymptotics for large areas and discuss the Bekenstein-Hawking area law. Finally we will look at the relevance of our results with regard to the  Schwarzschild quasi-normal modes link and show, in particular, that this connection is independent of the choice of $SU(2)$ or $SO(3)$ as the internal symmetry group.

The paper is structured as follows. After this introduction wee devote section \ref{abck} to review some important facts about the ABCK entropy definition. Section \ref{fluxoperator} deals with the flux-area operator that we propose to use here. In section \ref{sec4} we give some definitions and state the main result of the paper in theorem \ref{entropy}. We give the details of the state counting needed in the entropy computation in section \ref{counting}. A discussion of several relevant physical issues follows in section \ref{physicalissues}. The last section is devoted to conclusions and comments and we end the paper with several appendices where we prove some results used in the text and give some examples. Throughout the paper $\gamma$ stands for the Immirzi parameter and $\ell_P$ denotes the Planck length.

\section{The ABCK quantum horizon}{\label{abck}}

In this section we introduce and review some elements of the ABCK model of quantum
isolated horizons necessary to understand the definition and calculation of black hole entropy in LQG.
The starting point of the ABCK approach is a non-perturbative quantization based on a Hamiltonian framework. The arena for the Hamiltonian formalism is a 3-manifold $M$ homeomorphic to the complement of the unit open ball in $\mathbb{R}^3$. The boundary of $M$ is a 2-sphere, which we denote by $S$, that can be thought of as the intersection of an isolated horizon with $M$. As in \cite{ABK}, we will refer to $S$ simply as the `horizon'. A basic point of the ABCK approach is, precisely, to represent a black hole of a fixed classical area $a_\kappa$ as an isolated horizon that becomes an inner space-time boundary. Its presence requires the introduction of suitable surface terms in the gravitational action and implies the existence of bulk and surface gravitational degrees of freedom.  The consistent treatment of the latter leads to a Chern-Simons like quantization that requires the following area pre-quantization condition
\begin{equation}
a_\kappa\ =\ 4\pi\gamma\ell^2_{\rm P}\kappa,\ \ \ \ \ \kappa\in\mathbb{N}.
\label{a}\end{equation}

The horizon Hilbert space ${\cal H}_{\rm Hor}^\kappa$ is spanned by a
basis of quantum states
$$|(0)\rangle_{\rm Hor},\ldots ,|(b_1,\ldots,\,b_n)\rangle_{\rm Hor},\ldots$$
labeled by sequences  $(b_1,\ldots,\,b_n)$ of congruence classes
\begin{eqnarray*}
0\ \not =\ b_i\in\mathbb{Z}_\kappa\,,\ \ \ \ i=1,\ldots,n\,,\ \ \ \
n\in\mathbb{N}\,,
\end{eqnarray*}
such that
\begin{eqnarray*}
\sum_{i=1}^nb_i\ =\ 0\in \mathbb{Z}_\kappa.
\end{eqnarray*}
The length $n$ of a sequence is arbitrary and, hence, $n$ ranges over all the
natural numbers. In the previous basis $|(0)\rangle_{\rm Hor}$ denotes the state corresponding to a sequence with the single element $0\in \mathbb{Z}_\kappa$.
\medskip

The bulk Hilbert space ${\cal H}_{\rm Bul}$ is spanned by the states
\begin{equation}
|(0),\cdots\rangle_{\rm Bul},\ldots,
|(m_1,j_1,\ldots,m_n,j_n),\cdots\rangle_{\rm Bul}\,,\ldots\label{bulkstates}
\end{equation}
where the half integers $j_i\in\frac{1}{2}\mathbb{N}$ correspond to irreducible representations of
$SU(2)$ and
\begin{equation}
m_i \in \{-j_i,\,-j_i+1,\ldots,j_i-1,j_i\}\ \ \ {\rm for\ \ every} \ \ \
i=1,\ldots,n,\ \ \ n\in\mathbb{N}.\label{jm}
\end{equation}
The quantum numbers $m_i$ and $j_i$ represent the quantum degrees of
freedom of the bulk geometry close to the horizon, and ``$\cdots$'' in
the bulk state $|(m_1,j_1,\ldots,m_n,j_n),\cdots\rangle_{\rm Bul}$  stands for the other bulk degrees of freedom away from the horizon.

The horizon quantum area operator used in \cite{ABK} comes from the
standard area operator defined in the kinematical Hilbert space of
LQG for an arbitrary 2-surface
\cite{Rovelli,AL,Thiemann}. When applied to the horizon it extends to the
bulk Hilbert space ${\cal H}_{\rm Bul}$ as follows
\begin{eqnarray} \hat{a}^{\rm LQG}\,|(m_1,j_1,\ldots,m_n,j_n),\cdots\rangle_{\rm Bul} &=&
a^{\rm LQG}(j_1,\ldots,j_n)\,
|(m_1,j_1,\ldots,m_n,j_n),\cdots\rangle_{\rm
Bul}\,,\nonumber\\
a^{\rm LQG}(j_1,\ldots,j_n) &:=& 8\pi\gamma \ell^2_{P}\sum_{i=1}^n\sqrt{j_i(j_i+1)}\,.\label{LQGarea}
\end{eqnarray}

The total Hilbert space ${\cal H}^{\kappa}_{\rm Tot}$ of physical states
is a subspace of the the tensor product ${\cal H}^{\kappa}_{\rm Hor}\otimes {\cal
H}_{\rm Bul}$,
\begin{equation} {\cal H}^{\kappa}_{\rm Tot}\ <\ {\cal H}^{\kappa}_{\rm Hor}\otimes {\cal
H}_{\rm Bul}\,, \end{equation}
spanned by all the vectors of the form
\begin{equation}
|(b_1,\ldots,b_n)\rangle_{\rm Hor}\otimes
|(m_1,j_1,\ldots,m_n,j_n),\cdots\rangle_{\rm Bul},\ \ \ \  n\in \mathbb{N}_0\,,
\end{equation}
satisfying the following constraint
\begin{equation}\label{constraint}
b_i = -2m_i \quad (\mathrm{mod}\,\kappa)\,,\ \  {\rm for }\ \ i=1,\ldots,n\,.
\end{equation}

At this point the next step in the standard ABCK framework is to introduce an area interval $[a_\kappa-\delta,a_\kappa+\delta]$ with a $\delta$ of the order of $\ell_P^2$. The entropy is computed by first tracing out the bulk degrees of freedom to get a density matrix that describes a maximal entropy mixture of surface states with area eigenvalues in $[a_\kappa-\delta,a_\kappa+\delta]$. The value of the entropy is finally obtained by counting the number of allowed lists $(b_1,\ldots, b_n)$ of non-zero elements of $\mathbb{Z}_\kappa$ satisfying $b_1+\cdots+b_n=0$, such that $b_i=-2m_i\, (\mathrm{mod}\, \kappa)$ for some permissible third spin components $(m_1,\ldots,m_n)$. Here permissible means that there exists a list of non-vanishing spins $(j_1,\ldots, j_n)$ such that each $m_i$ is a spin component of $j_i$ and
\begin{eqnarray}
a_\kappa-\delta\leq a^{\rm LQG}(j_1,\ldots,j_n)=8\pi\gamma\ell_P^2\sum_{i=1}^n\sqrt{j_i(j_i+1)}\leq a_\kappa+\delta.\label{abk_LQG}
\end{eqnarray}
The counting of $b$-labels amounts to the determination of the dimension of the Hilbert subspace of ${\cal H}^{\kappa}_{\rm Hor}$ that represents the black hole degrees of freedom.

The introduction of a suitable area interval is crucial at this point because the pre-quantized area $a_\kappa$ \textit{does not belong} to the spectrum of the area operator $\hat{a}^{\mathrm{LQG}}$. In the ABCK model, however, there is another conceivable choice for the  quantum area operator --that we refer to as $\hat{a}^{\rm flux}$-- which classically corresponds to the same horizon area but has the property that the pre-quantized area $a_\kappa$ \textit{does belong} to its spectrum. As we will show in section \ref{fluxoperator}, the eigenvalues $a^{\rm flux}(m_1,\ldots,m_n)$ of $\hat{a}^{\rm flux}$ are labeled only by the spin components $(m_1,\ldots,m_n)$. As a consequence of this, the condition (\ref{abk_LQG}) of the standard ABCK framework can be consistently replaced by
\begin{eqnarray}
a_\kappa=a^{\rm flux}(m_1,\ldots,m_n)\,.\label{abk_flux}
\end{eqnarray}
This choice naturally leads to a definition of the black hole entropy that  requires tracing out the bulk degrees of freedom in a degenerate area interval $[a_\kappa-\delta,a_\kappa+\delta]$ with $\delta=0$. This introduces some simplifications in the resulting formalism and allows us to eliminate this arbitrariness associated to $\delta$. In the following sections, we discuss in detail $\hat{a}^{\rm flux}$ and explore the consequences of using it in the black hole entropy definition.

\section{The flux-area operator}{\label{fluxoperator}}
The use of an inner boundary $S$ to model the black hole horizon in the ABCK formalism carries an associated ambiguity in the definition of the quantum area operator for the horizon. This  ambiguity is more than the standard ordering one characteristic of quantum mechanics. It follows from the extra, non-dynamical,  structure defined on the horizon 2-surface $S$ in the ABCK model, namely a fixed Lie algebra $\mathfrak{su}(2)$ valued
function
$$r:S\rightarrow {\rm \mathfrak{su}(2)}.$$
Classically, the metric tensor of the 3-dimensional bulk manifold is
encoded in the Ashtekar vector density triad $\tilde{E}^a_I$ where
the indexes $a$ and $I$ respectively correspond to local
coordinates in the bulk, and the Lie algebra. The vector density
field $\tilde{E}^a_Ir^I$ (we use the scalar product $-2$Tr in $\mathfrak{su}(2)$)
is orthogonal to the horizon 2-sphere $S$,  and the flux of the
normal vector to the horizon can be written as\footnote{Given a
differential  $m$-form $\omega$ and a $m$-submanifold $M$, $\int_M
|\omega|$ is defined in the obvious way.}
\begin{equation}
a^{\rm flux}(\tilde{E},r)\ =\ \frac{1}{2}\int_S |\,\tilde{E}^a_Ir^I
\epsilon_{abc}\,\mathrm{d}x^b\wedge \mathrm{d}x^c\,|\,,
\end{equation}
where $x^a$ are local coordinates in the bulk, $\epsilon_{abc}$ is the alternating symbol, and the absolute value guarantees that the integrand is positive regardless the orientation of $S$. The quantum area-flux operator that we are proposing to use here is given by
\begin{equation}
\hat{a}^{\rm flux}\ =\ a^{\rm flux}(\widehat{\tilde{E}},r),
\end{equation}
where $r$ is left unquantized. This operator represents the quantum counterpart of the flux of $\tilde{E}^a_I r^I$ through the horizon, namely
\begin{equation} \hat{a}^{\rm flux}\,|(m_1,j_1,\ldots,m_n,j_n),\cdots\rangle_{\rm Bul}\ =\ a^{\rm flux}(m_1,\ldots,m_n)
|(m_1,j_1,\ldots,m_n,j_n),\cdots\rangle_{\rm Bul}\,,\label{fluxarea0}
\end{equation}
where
\begin{equation}
a^{\rm flux}(m_1,\ldots,m_n)\ =\ 8\pi\gamma \ell_{P}^2\sum_{i=1}^n |m_i|\,.\label{fluxarea}
\end{equation}
The spectrum of this operator, $sp(\hat{a}^{\rm flux})=4\pi\gamma\ell_{P}^2\mathbb{N}_0$,  is equidistant being $\Delta a=4\pi\gamma\ell_{P}^2$ the distance between two consecutive eigenvalues. Furthermore, and very important for our purposes, the pre-quantized values of the area $a_\kappa$ belong to $sp(\hat{a}^{\rm flux})$.

Notice that if the values of $r$ at each point of the surface
$S$ were not given (``gauge fixed'') by the ABCK model, one could try to alternatively
express it as a function $\rho(\tilde{E})$ of the triad field.
Substituting $r$ for $\rho(\tilde{E})$ in $a^{\rm
flux}(\tilde{E},r)$  would then give the standard area function
\begin{equation}
a(\tilde{E})\ =\ a^{\rm flux}(\tilde{E},\rho(\tilde{E})).
\end{equation}
On the classical phase space the functions $a^{\rm
flux}(\tilde{E},r)$ and $a^{\rm flux}(\tilde{E},\rho(\tilde{E}))$
coincide. However, the quantum area operator
\begin{equation}
\hat{a}^{\rm LQG}\ =\ a^{\rm
flux}(\widehat{\tilde{E}},\rho(\widehat{\tilde{E}}))
\end{equation}
is given by (\ref{LQGarea}) which differs from   the quantum flux-area
operator $a^{\rm flux}(\widehat{\tilde{E}},r)$ defined by
(\ref{fluxarea0}) and (\ref{fluxarea}).

It is important to point out that the advantage of the standard quantum area operator $\hat{a}^{\rm
LQG}$ is that it is available for \textit{arbitrary} 2-surfaces in the bulk.
Using this operator we can treat the intrinsic geometry of the horizon
on the same footing as the 2-geometry of any other 2-surface. This is the reason why that operator was used in the original ABCK model. However there are suggestive physical arguments that imply that the quantum black hole area should be
quantized and the spectrum should be equidistant\footnote{This has been suggested in \cite{BM}. As we will see in section  \ref{reactivation}, there are also some compelling arguments coming from the study of quasinormal modes pointing in the same direction.}. These arguments, together with the availability within LQG of a natural area operator $\hat{a}^{\rm flux}$ with the required equi-distant spectrum, lead us to study the consequences of using it to define the entropy. In fact, this operator has some extra advantages, in particular, as mentioned above the pre-quantized value of the
classical area (\ref{a}) belongs to the spectrum of $\hat{a}^{\rm
flux}$.  This is helpful because there is no need to introduce an interval of the form $[a_\kappa-\delta,a_\kappa+\delta]$ as in the original ABCK proposal in order to define the entropy. In our opinion this reinforces the non-trivial relationship between the quantum geometry in the bulk and the CS surface states originating from the quantum matching conditions. We want to point out, nonetheless, that the fact that we do not need to introduce \textit{a priori} an area interval at this stage does not mean that we should not use an interval in the definition of the entropy of a black hole; in fact, we will see that it is quite natural to do so in order to take into account, for example, the unavoidable finite resolution of measuring devices in any conceivable physical determination of a black hole entropy or define a quantum statistical microcanonical ensemble.

It is important to mention that some operators similar to the one proposed here have been considered in the literature by Krasnov and Sahlmann \cite{Krasnov,Hanno}. In these papers the authors have suggested to use area operators with eigenvalues defined in terms of the spins $j_i$. They appear quite naturally if one approximates the eigenvalues of the standard area operator (\ref{LQGarea}), for large spins $j_i$, according to
$$
\sqrt{j_i(j_i+1)}\sim j_i+\frac{1}{2}\sim j_i\,.
$$
This has led these authors to consider an area operator with the same spectrum as $\hat{a}^{\rm flux}$. Notice, however, that the eigenvalues of $\hat{a}^{\rm flux}$ involve the spin components $m_i$ whereas in these other proposals the action of the operators is defined in terms of the spins $j_i$. This changes the details of the state counting leading to the black hole entropy as will be commented in section \ref{comm}.

Finally, with the quantum flux-area operator, the relation of
the ABCK model with the black hole quasi-normal modes \cite{Dreyer},
which was destroyed in \cite{DL}, can be proven again, and will hold
even stronger than before: see section \ref{reactivation}.

\section{The quantum entropy and the flux-area operator}{\label{sec4}}

In this section we will give the precise entropy definition that we will use throughout the paper and enunciate the main result of this work. We start by giving some definitions that will be used in the following.

\begin{df}{\label{msequ}} Let $\kappa\in \mathbb{N}$ be a fixed value corresponding to the level of the CS-theory that gives rise to the prequantized area $a_\kappa=4\pi\gamma\ell^2_P \kappa$. We say the list $m=(m_1,\ldots,m_n)$ of  half-integers is $\kappa$-permissible list of spin components if it satisfies the area condition
$$
\sum_{i=1}^n |m_i|=\kappa/2\,.
$$
\end{df}
\begin{df}
Given a list $b=(b_1,\ldots, b_n)$ of non-zero elements of $\mathbb{Z}_\kappa$, we say it is $\kappa$-permissible if the following two conditions are satisfied
\begin{itemize}
\item[c1)] $b_1+\cdots+b_n=0\in \mathbb{Z}_\kappa$\,,
\item[c2)] $b_i= -2m_i\, (\mathrm{mod}\, \kappa)$ for some $\kappa$-permissible list of spin components $m=(m_1, . . . ,m_n)$.
\end{itemize}
We will denote by $\mathcal{B}_\kappa$ the set of all  $\kappa$-permissible $b$-lists of non-zero elements of $\mathbb{Z}_\kappa$. In the following the equality $\mathrm{mod}\, \kappa$ will be denoted with the symbol $\equiv$. Hence the condition c2) above will be written as $b_i\equiv -2m_i$.
\end{df}

The direct translation of the procedure given by ABCK  to define the entropy to the case when the $\hat{a}^{\rm flux}$ operator is used instead of $\hat{a}^{\rm LQG}$ is the following:
\begin{df}[black hole entropy]{\label{def_entropy}}
The entropy $S_{\rm bh}(a_\kappa)$ of a quantum horizon of area $a_\kappa=4\pi\gamma\ell_P^2\kappa$,
with $\kappa\in\mathbb{N}$, is given by the formula
$$S_{\rm bh}(a_\kappa) = \log \big(|\mathcal{B}_\kappa|+1\big)
\,,$$
where the 1 above comes from the trivial sequence.
\end{df}

Several comments are in order now. First notice that after the substitution of $\hat{a}^{\rm LQG}$ for $\hat{a}^{\rm flux}$ the labels $(j_1,\ldots,j_n)$ completely disappear from the definition of the entropy; in particular there is no need to define permissible lists of $j$'s although one can trivially obtain one by considering $(|m_1|,\ldots,|m_n|)$ for a given list of $m$'s satisfying $a^{\rm flux}(m_1,\ldots,m_n)=a_\kappa\,.$ Also notice that the interval has disappeared from the definition and has been substituted by a sharp equality condition. Finally remember that we are counting horizon states. In this regard we want to point out that we will implement in the next section the philosophy of \cite{DL} of changing this counting problem into an equivalent one involving only the counting of third spin components ($m$-labels) with the only difference that we will not need to introduce any type of area interval as in \cite{DL}.

The central result of this paper is the following exact, closed form, expression for the entropy of a black hole obtained according to the previous definition.

\bigskip

\begin{thr}{\label{entropy}} Let $\kappa\in \mathbb{N}$ the CS-level and $a_\kappa=4\pi\gamma\ell^2_P\kappa$. For even values of the CS-level the entropy of a quantum horizon of  area $a_{2u}$, $u\in\mathbb{N}$, is given by
\begin{eqnarray*}
S_{\rm bh}(a_{2u})&=&\log\Bigg(\frac{(-1)^u\sqrt{\pi}}{4\Gamma(3/2-u)\cdot u! }\,_{2}F_1(1/2,-u;3/2-u;9)+2^{2u}-2^{u-1}(u+3)-2u+2\Bigg)\,,
\end{eqnarray*}
where $\Gamma$ and $_{2}F_{1}$ are, respectively, the Gamma function and the Gauss's hypergeometric function. On the other hand,  for odd values of the CS-level the entropy of a quantum horizon of area $a_{2u+1}$,  $u\in\mathbb{N}_0$, is
$$
S_{\rm bh}(a_{2u+1})=\log\Big(2^{2u+1}-2u-1\Big)\,.
$$
\end{thr}

\bigskip

The proof of this result is the scope of the next section.

\section{The quantum entropy counting}{\label{counting}}
In order to proof theorem \ref{entropy} we have to count states according to the definition \ref{def_entropy}; this amounts to counting the elements of $\mathcal{B}_\kappa$. To this end we will follow a strategy close to the one employed in \cite{DL} and translate the problem into one involving only sequences $m=(m_1,\ldots,m_n)$ of spin components.  In the following we will find it convenient to work with integer sequences  $\nu=2m$ so we give the following auxiliary definition.
\begin{df}[integer permissible lists]{\label{nuseq}}
We say that the list $\nu=(\nu_1,\ldots,\nu_n)\in \mathbb{Z}^n$ is a $\kappa$-permissible list of integers if
$$
\sum_{i=1}^n |\nu_i|=\kappa\,.
$$
We will denote by $\mathcal{N}_\kappa$ the set of all $\kappa$-permissible list of integers.
\end{df}
The set $\mathcal{N}_\kappa$ is bijective with the set of $\kappa$-permissible list of spin components  introduced in definition \ref{msequ}, and the set of all $\kappa$-permissible $b$-sequences can be written as
\begin{eqnarray*}
\mathcal{B}_{\kappa}&=&\Big\{b\,:\, \exists n\in \mathbb{N}\,,\, b=(b_1,\ldots,b_n)\in \mathbb{Z}_\kappa^n\,, \,\sum_{i=1}^nb_i= 0\,,\,0\neq b_i\equiv -\nu_i\,,\,(\nu_1,\ldots,\nu_n)\in \mathcal{N}_\kappa\Big\}\,.
\end{eqnarray*}
Notice that the condition $0\neq b_i\equiv -\nu_i$ implies $\nu_i\neq 0$ and  also $\nu_i\neq \pm \kappa$ (the other integer multiples of $\kappa$ are trivially excluded by the area condition $\sum_{i=1}^n|\nu_i|=\kappa$ appearing in definition \ref{nuseq}). Moreover $\sum_{i=1}^nb_i=0$ implies $\sum_{i=1}^n\nu_i\equiv 0$. Then, using  $\sum_{i=1}^n\nu_i=y\kappa$, $y\in\mathbb{Z}$, we have $\kappa=\sum_{i=1}^n|\nu_i|\geq |\sum_{i=1}^n\nu_i|=|y|\kappa$, so $y=0, \pm 1$. Hence, it suffices to restrict ourselves to the following subset of the $\kappa$-permissible list of integers
\begin{eqnarray*}
\mathcal{V}_\kappa&=&\Big\{\nu\,:\, \exists n\in \mathbb{N}\,,\, \nu=(\nu_1,\ldots,\nu_n)\in \mathbb{Z}_*^n\,,\, |\nu_i|\neq \kappa\,,\, \sum_{i=1}^n|\nu_i|= \kappa,\,  \sum_{i=1}^n \nu_i=\begin{cases}0,\textrm{ or }\\
\pm \kappa\end{cases}\Big\}\,,
\end{eqnarray*}
where $\mathbb{Z}_*=\mathbb{Z}\setminus\{0\}$. Notice that $\mathcal{V}_\kappa\subset \mathcal{N}_\kappa$. The set $\mathcal{V}_\kappa$ can be partitioned as the disjoint union
$$
\mathcal{V}_\kappa= \mathcal{V}^+_\kappa\cup \mathcal{V}^0_\kappa\cup \mathcal{V}^-_\kappa\,,
$$
where
\begin{eqnarray*}
\mathcal{V}^0_\kappa&=&\{\nu\,:\, \exists n\in \mathbb{N}\,,\, \nu=(\nu_1,\ldots,\nu_n)\in \mathbb{Z}_*^n\,,\, |\nu_i|\neq \kappa\,,\,\, \sum_{i=1}^n|\nu_i|= \kappa, \sum_{i=1}^n \nu_i=0\}\,,\\
\mathcal{V}^{\pm}_\kappa&=&\{\nu\,:\, \exists n\in \mathbb{N}\,,\, \nu=(\nu_1,\ldots,\nu_n)\in \mathbb{Z}_*^n\,,\, |\nu_i|\neq \kappa\,,\,\, \sum_{i=1}^n|\nu_i|= \kappa, \sum_{i=1}^n \nu_i=\pm \kappa\}.
\end{eqnarray*}
These sets can be given also in the following equivalent but simpler forms
\begin{eqnarray}
\mathcal{V}^0_\kappa&=&\{\nu\,:\, \exists n\in \mathbb{N}\,,\, \nu=(\nu_1,\ldots,\nu_n)\in \mathbb{Z}_*^n\,,\, \sum_{i=1}^n|\nu_i|= \kappa, \sum_{i=1}^n \nu_i=0\}\,,\label{nu0}\\
\mathcal{V}^{\pm}_\kappa&=&\{\nu\,:\, \exists n>2\,,\, \nu=(\nu_1,\ldots,\nu_n)\in \mathbb{Z}_*^n\,,\, \sum_{i=1}^n|\nu_i|= \kappa, \sum_{i=1}^n \nu_i=\pm \kappa\}\,.\label{nupm}
\end{eqnarray}
First notice that we do not need to explicitly include the condition $|\nu_i|\neq \kappa$ in the definition of $\mathcal{V}^0_\kappa$ because if we had some $\nu_i=\kappa$ the condition $\sum_{i=1}^n|\nu_i|= \kappa$ would tell us that there can be only one such $\nu_i$ and hence it would be impossible to have $\sum_{i=1}^n \nu_i=0$. This implies, by the way, that there are no sequences of unit length in $\mathcal{V}^0_\kappa$.
Each of the sets $\mathcal{V}^{\pm}_\kappa$ can be split as one containing unit length sequences and another containing the rest. Unit length sequences are excluded by the impossibility of simultaneously satisfying the conditions $|\nu_1|\neq \kappa$ and $\sum_{i=1}^n \nu_i=\nu_1=\pm \kappa$. Finally, for sequences of length greater or equal than 2, the condition $\sum_{i=1}^n|\nu_i|= \kappa$ with non-zero $\nu_i$ directly excludes that any of them is $\pm\kappa$. After these considerations we see that it is possible to write $\mathcal{B}_\kappa$ in terms of $\mathcal{V}_\kappa$ as
\begin{eqnarray*}
\mathcal{B}_{\kappa}&=&\Big\{b\,:\, \exists n\in \mathbb{N}\,,\, b=(b_1,\ldots,b_n)\in \mathbb{Z}_\kappa^n\,, \,\sum_{i=1}^nb_i= 0\,,\, 0\neq  b_i\equiv -\nu_i\,, \,(\nu_1,\ldots,\nu_n)\in \mathcal{V}_\kappa\Big\}\,.
\end{eqnarray*}
It is straightforward to prove now the following result.
\begin{prop}
The equality modulo $\kappa$ defines an equivalence relation on $\mathcal{V}_\kappa$, that we also denote by $\equiv$, as follows: $\nu$ and $\nu^\prime$ are equivalent ($\nu\equiv \nu'$) if and only if  $\nu$ and $\nu'$ have the same length and $\nu_i\equiv \nu'_i$ for all $i$. Hence, by using that for any $\nu\in \mathcal{V}_\kappa$ the sequence $b\equiv -\nu$ always belongs to $\mathcal{B}_\kappa$,  the quotient space $\mathcal{V}_\kappa/\equiv$ can be identified with $\mathcal{B}_{\kappa}$.
\end{prop}
Notice that, if all the equivalence classes of $\mathcal{V}_\kappa/\equiv$ had only one element there would exist a bijection between $\mathcal{B}_\kappa$ and $\mathcal{V}_\kappa$ and our counting problem would boil down to the simpler one of counting the elements of $\mathcal{V}_\kappa$. However, there is an overcounting issue if we only look at $\mathcal{V}_\kappa$ owing to the existence of equivalence classes with more that one element. In other words, there are different $\nu$-sequences corresponding to the same $b$-sequence such that $\nu=(\kappa-1,1)$ and $\nu^\prime=(-1,1-\kappa)$. This forces us to carefully study $\mathcal{V}_\kappa$.

In order to proceed now it is convenient to consider odd and even values of $\kappa$ separately.

\subsection{Prequantized odd numbers}

In the case of odd values of $\kappa=2u+1$ the strategy to deal with the overcounting issue is straightforward. We first show that $\mathcal{V}^0_{2u+1}=\emptyset$. Hence, we only have to consider $\mathcal{V}^{\pm}_{2u+1}$. It is easy to prove that there is no possible overcounting within each of these sets so that the only remaining possibility for overcounting is that two sequences belonging respectively to $\mathcal{V}^{+}_{2u+1}$ and $\mathcal{V}^{-}_{2u+1}$ give the same $b$-sequence. These cases are very easy to characterize and count: they are the length-two sequences in $\mathcal{V}^{\pm}_{2u+1}$.

In order to complete this program we need to prove several propositions. In the first, by using that the parity of the sum of the elements of the sequences $(\nu_1,\ldots,\nu_n)$ and $(|\nu_1|,\ldots,|\nu_n|)$ is the same, we prove that $\mathcal{V}^0_{2u+1}=\emptyset$.

\begin{prop}
There is no sequence $\nu=(\nu_1,\ldots,\nu_n)$ such that
$$
\sum_{i=1}^n \nu_i=0 \quad \textrm{ and } \quad \sum_{i=1}^n|\nu_i|=2u+1= 1\,(\mathrm{mod}\,2)\,,
$$
and, hence, $\mathcal{V}^0_{2u+1}=\emptyset\,.$
\end{prop}
\begin{proof}
Using $\nu_i= |\nu_i|\,(\mathrm{mod}\,2)$ we get
$$
0=\sum_{i=1}^n \nu_i= \sum_{i=1}^n|\nu_i|= 1\,(\mathrm{mod}\,2)
$$
which is imposible.
\end{proof}
This means that
$$
\mathcal{V}_{2u+1}=\mathcal{V}^+_{2u+1}\cup \mathcal{V}^-_{2u+1}
$$
Notice that if $\nu\in \mathcal{V}^+_\kappa$ then $0<\nu_i=|\nu_i|<\kappa$ for all $i$. This trivially follows from the fact that in $\mathcal{V}^{+}_\kappa$ the equality $\sum_{i}|\nu_i|=\sum_i\nu_i=\kappa$ holds. Also, if $\nu^\prime\in \mathcal{V}^-_\kappa$ then $-\kappa<\nu_i^\prime=-|\nu_i^\prime|<0$ for all $i$.
It is then clear that, if $\nu,\nu'\in\mathcal{V}^+_{2u+1}$, the condition $\nu'\equiv \nu$ implies $\nu=\nu'$. This also true in $\mathcal{V}^-_{2u+1}$. This proves that there is no danger of overcounting within each of the sets $\mathcal{V}^+_\kappa$ and $\mathcal{V}^-_\kappa$, i.e. each equivalence class on $\mathcal{V}_\kappa/\equiv$ has, at most, one element of $\mathcal{V}^+_\kappa$ and the same is true for $\mathcal{V}^-_\kappa$. The next proposition identifies the sequences in  $\mathcal{V}^+_\kappa$ and $\mathcal{V}^-_\kappa$ leading to the same $b$-sequence.

\begin{prop}\label{propV3} Let $\nu\in\mathcal{V}^+_{2u+1}$. Then there exists a sequence $\nu'\in\mathcal{V}^-_{2u+1}$ satisfying $\nu'\equiv \nu$ if and only if $\nu$ has two components $\nu=(\nu_1,\nu_2)$ and $\nu'=(\nu_1-2u-1,\nu_2-2u-1)$. There are $2u$ such sequences in $\mathcal{V}^+_{2u+1}$. Therefore, the only equivalence classes on $\mathcal{V}_\kappa/\equiv$ with more than one element are those of the form $\{(\nu_1,\nu_2),(\nu_1',\nu_2')\}$ where $(\nu_1,\nu_2)\in\mathcal{V}^+_{\kappa}$ and $(\nu'_1,\nu'_2)=(\nu_1-\kappa,\nu_2-\kappa)\in\mathcal{V}^-_\kappa$.
\end{prop}
\begin{proof}
The condition is sufficient. If $\nu=(\nu_1,\nu_2)\in\mathcal{V}^+_\kappa$ then $0<\nu_i<\kappa$ and
$$
\nu_1+\nu_2=\kappa.
$$
The sequence $\nu'=(\nu_1',\nu_2')$, with $-\kappa<\nu_1'=\nu_1-\kappa<0$, $-\kappa<\nu_2'=\nu_2-\kappa<0$, satisfies
$$
\nu'_1+\nu'_2=-\kappa\,,\quad |\nu_1'|+|\nu_2'|=\kappa\,,
$$
and hence $\nu'\in\mathcal{V}^-_{\kappa}$.

Trivially, the condition is also necessary. Let $\nu_i'\equiv \nu_i \,$ for all $i$. Then $\nu_i'=\nu_{i}-\kappa s_i$, with $s_i=0$ or 1. So we have
$$
-\kappa=\sum_{i=1}^n\nu_i'=\sum_{i=1}^n\nu_i-\kappa\sum_{i=1}^ns_i=\kappa-\kappa\sum_{i=1}^ns_i\Rightarrow \exists i_1, i_2\,:\, s_{i_1}=s_{i_2}=1\,,\, s_i=0\,,\, i\neq i_1,i_2.
$$
but $s_i=0$ implies $\nu_i'=\nu_i>0$ which is imposible in $\mathcal{V}^-_\kappa$. Hence $\nu$ has only two components. It is straightforward to see that there are $2u$ such sequences in each $\mathcal{V}^{\pm}_{2u+1}$.
\end{proof}

\noindent Finally, as shown in appendix \ref{AppPartitions}, it is straightforward to find that
$$
|\mathcal{V}^+_{2u+1}|=|\mathcal{V}^-_{2u+1}|=2^{2u}-1.
$$
Hence, as it is claimed in theorem \ref{entropy}, for black holes with odd $\kappa=2u+1$, the entropy is
$$
S_{\rm bh}(a_{2u+1})=\log\Big(|\mathcal{B}_{2u+1}|+1\Big)=\log\Big(2|\mathcal{V}^+_{2u+1}|-2u+1\Big)=\log\Big(2^{2u+1}-2u-1\Big)\,.
$$

\subsection{Prequantized even numbers}

Our strategy now to deal with the overcounting problem when $\kappa=2u$, $u\in\mathbb{N}$, is the following:

\bigskip

\noindent(1) Show that if $\nu\in \mathcal{V}^+_{2u}$ and $\nu'\in \mathcal{V}^+_{2u}$, then $\nu\equiv \nu'$ is equivalent to $\nu=\nu'$.

\noindent(2) Show that if $\nu\in \mathcal{V}^-_{2u}$ and $\nu'\in \mathcal{V}^-_{2u}$, then $\nu\equiv \nu'$ is equivalent to $\nu=\nu'$.

\noindent(3) Characterize the equivalent sequences in $\mathcal{V}^0_{2u}$. They reduce to $(u,-u)$ and $(-u,u)$

\noindent(4) Show that if $\nu\in \mathcal{V}^+_{2u}$ and $\nu'\in \mathcal{V}^-_{2u}$, then $\nu\equiv \nu'$ is equivalent to $\nu=(\nu_1,\nu_2)$ and, in addition, $\nu'=(\nu_1-2u,\nu_2-2u)$.

\noindent(5) Show that if $\nu\in \mathcal{V}^\pm_{2u}$ and $\nu'\in \mathcal{V}^0_{2u}$, then $\nu\equiv \nu'$ is equivalent to the existence of an index $i_0$ such that $\nu_{i_0}=\pm u$, $\nu'_{i_0}=\mp u$, and $\nu_i=\nu_i'$, whenever $i\neq i_0$.

\bigskip

Points (1), (2) are trivial as shown above. Point (4) is actually proved in proposition \ref{propV3} because the parity of $\kappa$ plays no role in its proof.  We have then to concentrate only on points (3) and (5).

Let us first consider point (3) consisting in the characterization of equivalent sequences in $\mathcal{V}^0_{2u}$. To this end we need to prove the following propositions.

\begin{prop}{\label{cota}}
If $\nu\in \mathcal{V}^0_{2u}$ then $|\nu_i|\leq u$.\end{prop}

\begin{proof}Otherwise, suppose that $|\nu_{i_0}|>u$, then $\sum_{i\neq i_0}\nu_i=-\nu_{i_0}$, and hence $2u=\sum_{i}|\nu_i|=|\nu_{i_0}|+\sum_{i\neq i_0}|\nu_i|>u+\sum_{i\neq i_0}|\nu_i|\geq u+|\sum_{i\neq i_0}\nu_i|>u+u=2u$ which is impossible.
\end{proof}

\begin{prop}{\label{p0}} Let $\nu,\nu^\prime\in \mathcal{V}^0_{2u}$. If $\nu'\equiv \nu$ but $\nu'\neq \nu$ then either $\nu=(u,-u)$ and $\nu'=(-u,u)$ or $\nu=(-u,u)$ and $\nu'=(u,-u)$.
\end{prop}

\begin{proof} The equivalence of $\nu$ and $\nu'$ implies that $\nu_i'=\nu_i+2s_i u$, for all $i$, with $s_i\in\mathbb{Z}$. The fact that $|\nu_i|\leq u$ and $|\nu_i'|\leq u$ actually restricts the values of the $s_i$ to be $0,\pm1$. As we have that $\sum_{i=1}^n\nu_i=\sum_{i=1}^n\nu_i'=0$ we must have as many values of the $s_i$ equal to $+1$ as values equal to $-1$. The condition $\nu\neq\nu'$ means that at least one value of $s_i$, say $s_{i_1}$, is equal to $+1$. We have then
$$
\nu'_{i_1}=\nu_{i_1}+2u
$$
which implies that $\nu_{i_1}=-|\nu_{i_1}|$. Finally we have
$$
u\geq|\nu'_{i_1}|=2u-|\nu_{i_1}|\geq u
$$
and, hence $\nu'_{i_1}=u$ and $\nu_{i_1}=-u$.

As positive and negative values of $s_i$ come in pairs there must be another $s_i$, say $s_{i_2}$, equal to $-1$, the same argument tells us that  $\nu'_{i_2}=-u$ and $\nu_{i_2}=u$. By noting that
$$
\sum_{i=1}^n|\nu_i|=\sum_{i=1}^n|\nu'_i|=2u
$$
the sequences $\nu$ and $\nu'$ can only have two elements and hence they are either $\nu=(u,-u)$ and $\nu'=(-u,u)$ or $\nu=(-u,u)$ and $\nu'=(u,-u)$.
\end{proof}

The last proposition tells us that there is only one equivalence class in $\mathcal{V}_\kappa/\equiv$ with more than one representative in $\mathcal{V}_\kappa^0$. This class simultaneously contains $(-u,u)$ and $(u,-u)$ and the rest of the representatives belong to $\mathcal{V}_\kappa^\pm$ [actually they are $(u,u)\in\mathcal{V}_\kappa^+$ and $(-u,-u)\in\mathcal{V}_\kappa^-$, as will be shown bellow].

\bigskip

We deal now with point (5). To this end we need to prove the following result.

\begin{prop}{\label{p1}} Let  $\nu\in \mathcal{V}^+_{2u}$ and  $\nu'\in \mathcal{V}^0_{2u}$. The condition $\nu'\equiv \nu$ is equivalent to the existence of an index $i_0$ such that $\nu_{i_0}=u=-\nu'_{i_0}$ and $\nu_{i}=\nu'_{i}$ for $i\neq i_0$.
\end{prop}
\begin{proof} Clearly if $\nu\in \mathcal{V}^+_{2u}$ and  $\nu'\in \mathcal{V}^0_{2u}$ satisfying  $\nu_{i_0}=u=-\nu'_{i_0}$ and $\nu_{i}=\nu'_{i}$, for $i\neq i_0$, then $\nu'\equiv \nu$.
On the other hand, if $\nu\in \mathcal{V}^+_{2u}$ and  $\nu'\in \mathcal{V}^0_{2u}$ and $\nu'\equiv \nu$ then $\nu_i'=\nu_i-2s_iu$, $s_i=0$ or $1$ (remember that $2u>\nu_i>0$ and, hence, the values of $s_i$ different from 0 or 1 are excluded). Hence
$$
0=\sum_{i=1}^n\nu'_i=\sum_{i=1}^n \nu_i-2u\sum_{i=1}^ns_i=2u-2u\sum_{i=1}^ns_i\,.
$$
We see then that there exists just a single index $i_0$ such that  $s_{i_0}=1$ and the remaining $s_i$ vanish ($s_i=0\,, i\neq i_0$). If $\nu_{i_0}>u$ we would have $|\nu_{i_0}'|=2u-|\nu_{i_0}|<u<|\nu_{i_0}|$ and hence
$$
\sum_{i=1}^n |\nu_i'|<\sum_{i=1}^n |\nu_i|=2u
$$
which is not allowed. Similarly, if $\nu_{i_0}<u$ we would have $|\nu_{i_0}'|=2u-|\nu_{i_0}|>u$ which, in view of proposition \ref{cota}, is not allowed. Therefore we conclude $\nu_{i_0}=u$.

\end{proof}

A similar argument proves

\begin{prop}{\label{p2}} Let  $\nu\in \mathcal{V}^-_\kappa$ and  $\nu'\in \mathcal{V}^0_\kappa$. The condition $\nu'\equiv \nu$ is equivalent to the existence of an index $i_0$ such that $\nu_{i_0}=-u=-\nu'_{i_0}$ and $\nu_{i}=\nu'_{i}$ for $i\neq i_0$.
\end{prop}

As a direct consequence of propositions \ref{p1} and \ref{p2}, the equivalence classes of $\mathcal{V}_{2u}/\equiv$ that do not have representatives on $\mathcal{V}_{2u}^0$ are those with representatives belonging to the sets $$\tilde{\mathcal{V}}^\pm_{2u}=\{\nu\in\mathcal{V}^\pm_{2u}\,:\, \nu_i\neq u\,,\, \forall i\}\subset \mathcal{V}^\pm_{2u}\,.$$
Also, as we have shown in point (5), two different elements, one in $\tilde{\mathcal{V}}^+_{2u}$ and the other in $\tilde{\mathcal{V}}^-_{2u}$, do not always represent different equivalence classes. The overcounting is precisely due to the sequences of length two that appear in proposition \ref{propV3}. There are $2u-2$ sequences with two components in $\tilde{\mathcal{V}}^+_{2u}$. Finally, the overcounting of equivalence classes in $\mathcal{V}_{2u}^0$ is only associated with the sequences $(u,-u)$ and $(-u,u)$ as shown in proposition \ref{p0}. Hence, we have proved the following:

\begin{prop} The cardinality of $\mathcal{B}_{2u}$ coincides with the cardinality of the quotient space $\mathcal{V}_{2u}/\equiv$ \Blue{and} is given by the following expression
$$
|\mathcal{B}_{2u}|=|\mathcal{V}_{2u}^0|-1+2|\tilde{\mathcal{V}}^+_{2u}|-(2u-2)
=|\mathcal{V}_{2u}^0|+2|\tilde{\mathcal{V}}^+_{2u}|-2u+1\,.
$$
\end{prop}

\bigskip

Finally (see appendices \ref{AppPartitions} and \ref{AppGen}), by using
$$
|\mathcal{V}_{2u}^0|= \frac{(-1)^u \sqrt{\pi}}{4 \Gamma(3/2 - u) u!}\, _{2}F_{1}(1/2, -u; 3/2 - u; 9)\,,\quad |\tilde{\mathcal{V}}^\pm_{2u}|=2^{2u-1}-2^{u-2}(u+3)\,,
$$
we have
$$
|\mathcal{B}_{2u}|=\frac{(-1)^u \sqrt{\pi}}{4 \Gamma(3/2 - u)\cdot u!}\, _{2}F_{1}(1/2, -u; 3/2 - u; 9)+2^{2u}-2^{u-1}(u+3)-2u+1\,.
$$
Hence, as claimed in theorem \ref{entropy}, the entropy for black holes with even $\kappa=2u$ is
\begin{eqnarray*}
S_{\rm bh}(a_{2u})&=&\log\Big(|\mathcal{B}_{2u}|+1\Big)\\
&=&\log\Bigg(\frac{(-1)^u\sqrt{\pi}}{4\Gamma(3/2-u)\cdot u! }\,_{2}F_1(1/2,-u;3/2-u;9)+2^{2u}-2^{u-1}(u+3)-2u+2\Bigg)\,.
\end{eqnarray*}

\section{Physical issues}{\label{physicalissues}}

In this section we will discuss some physical issues in relation with our new proposal, in particular the Bekenstein-Hawking area law and the link with the Schwarzschild quasi-normal modes.

\subsection{The Bekenstein-Hawking area law}

The asymptotic behavior for large areas of the black hole entropy in the scheme that we are using in the paper can be computed in a straightforward way from the expressions appearing in theorem \ref{entropy}. The relevant formulas are
\begin{eqnarray*}
S_{\rm bh}(a_{2u+1})&=&\log\Big(2^{2u+1}-2u-1\Big)\,,\\
S_{\rm bh}(a_{2u})&=&\log\Bigg(\frac{(-1)^u\sqrt{\pi}}{4\Gamma(3/2-u)\cdot u! }\,_{2}F_1(1/2,-u,3/2-u,9)+2^{2u}-2^{u-1}(u+3)-2u+2\Bigg)\,.
\end{eqnarray*}
Hence, by using the asymptotic expansion for the hypergeometric function obtained in appendix \ref{AppAsym}, we get in the large $u$ regime
\begin{eqnarray*}
S_{\rm bh}(a_{2u+1})&\sim& \frac{a_{2u+1}\log 2}{4\pi\gamma\ell^2_P}\,,\\
S_{\rm bh}(a_{2u})&\sim&\log\Bigg( 3^{2u-1}\sqrt{\frac{2}{\pi u}}\Bigg)\sim \frac{a_{2u}\log 3}{4\pi\gamma\ell^2_P}-\frac{1}{2}\log \left(\frac{a_{2u}}{4\pi\gamma\ell^2_P}\right)\,.
\end{eqnarray*}

The first step in the analysis of our result is a comparison with the asymptotic form of the Bekenstein-Hawking area law,
$$
S_{\rm BH}(a)\sim \frac{a}{4\ell^2_P}\,, \quad a\rightarrow \infty,
$$
where $S_{\rm BH}$ is the Bekenstein-Hawking entropy. The law seems to be only strictly valid if we restrict our entropy $S_{\rm bh}$ only to``even areas'' $a_{2u}$ or ``odd areas'' $a_{2u+1}$.
However, the Bekenstein-Hawking law should be considered at a suitable scale,
larger the one corresponding to quantum gravity. The scale of our LQG calculation is given by the area gap $4\pi\gamma \ell^2_P$. Therefore, to make contact with the Bekenstein-Hawking law we should coarse grain our exact quantum entropy $S_{\rm bh}$ over an interval
$[a-\Delta, a+\Delta]$ centered around a given classical area $a$, where certainly
$$\Delta\ >\  4\pi\gamma \ell^2_P.$$
Another way to understand this would be by invoking the necessity to take into account the unavoidable finite resolution of any conceivable measuring device\footnote{The area gap relevant in our computations is $4\pi\gamma \ell^2_P$ so the resolution of the physical measuring apparatus should be quantified in comparison with this scale.} or the definition of a suitable quantum microcanonical ensemble.
In any case, given an interval $[a-\Delta, a+\Delta]$ of width $2\Delta>8\pi\gamma \ell^2_P$, we define $S^{(\Delta)}_{\rm BH}(a)$ to be the logarithm of the total number of the CS states which correspond to the quantum areas $a_\kappa\in [a-\Delta, a+\Delta]$, that is
\begin{equation}
S^{(\Delta)}_{\rm BH}(a)\ :=\ \log \left( \sum_{\kappa:a_\kappa\in[a-\Delta,a+\Delta]} e^{S_{\rm bh}(a_\kappa)}
\right).
\end{equation}
The expansion for large values of $a$ is
\begin{equation}
S_{\rm BH}^{(\Delta)}(a)\ \sim\ \frac{a\log 3}{4\pi\gamma\ell^2_P}-\frac{1}{2}\log \left(\frac{a}{4\pi\gamma\ell^2_P}\right).
\end{equation}
Notice that the dependence on the interval width $2\Delta$ drops out.

Now, the value of the Immirzi parameter can be fixed such that the Bekenstein-Hawking holds, namely
\begin{equation}
\gamma\ =\ \frac{\log 3}{\pi}\,.
\label{gamma}
\end{equation}

\subsection{The area gap}

Given this value of $\gamma$ we can calculate the distance between eigenvalues of the flux-area spectrum. The spacing between consecutive area eigenvalues is
\begin{equation}
\Delta a\ =\ 4\pi\gamma \ell_P^2\ =\ 4 \log 3
\,\ell_P^2 \label{4log3}.
\end{equation}
This value of the area gap is known for its very special properties \cite{Baez}. We describe them in the subsection \ref{reactivation}. But before that, several remarks are in order.

\subsection{The relevance of the even and odd sectors}

The microscopic behavior of the exact quantum entropy $S_{\rm bh}$ is reminiscent of the substructure found in \cite{val1,prlnos} for the entropy of black holes when the standard form of the area spectrum is considered. The even areas $a_{2u}$ are responsible for the large scale properties of the entropy, for the value of $\gamma$ given by (\ref{gamma}), and the logarithmic correction to the entropy with the usual $-1/2$ coefficient.
If there were any physical reasons to restrict the quantum areas to the even $\kappa$,
the only difference would be a doubling of the area gap (\ref{4log3}).

On the other hand, if there were any physical conditions requiring us to restrict ourselves to the  odd area subsector, we would find that consistency with the Bekenstein-Hawking law would require
$$\gamma_{\rm odd}\ =\ \frac{\log 2}{\pi}\,,$$
and in the consequence the area gap would have to be
$$\Delta a_{\rm odd}\ =\  8 \log 2\,\ell_P^2,$$
a value favored by the arguments presented in \cite{BM} that suggest that the spacing between consecutive allowed horizon areas should be of the form $4N\log2$ with $N\in\mathbb{N}$. In this case no logarithmic term would show up in the large area expansion.

In fact, the parity of the CS level $\kappa$ has a clear physical interpretation. The bulk states are constructed from embedded graphs. The numbers $(m_1,j_1,\ldots,m_n,j_n)$
introduced in section \ref{abck} correspond to the intersection of the graph with the surface $S$ of the horizon. Suppose the intersection splits the graph into two disjoint parts (contained ``inside''/``outside'' the black hole). This is a justified assumption. If there are no fermions in the theory, then each of the disjoint parts of the graph intertwines the tensor product
of the $SU(2)$ representations corresponding to the spins $j_1,\ldots,j_n$ into the trivial representation. Therefore, in that fermionless case, the extra necessary condition is
$$ \sum_{i=1}^n{j_i}\ \in\ \mathbb{N}. $$
It is equivalent to
$$ \frac{\kappa}{2}\ =\ \sum_{i=1}^n|m_i|\ \in\mathbb{N}. $$
In the general case, the vertices of the graph are coloured with fermions. Then, the $SU(2)$ representations corresponding to the numbers $j_1,\ldots,j_n$ are intertwined by the interior/exterior part of the graph into a representation defined in the tensor product of the spinor spaces of  the vertices contained inside/outside the black hole. Therefore, $\kappa$ is odd, provided the total spin of the fermions inside/outside the black hole is half-integer.

\subsection{Reactivation of the quasi-normal modes link}{\label{reactivation}}
As we have mentioned above the value (\ref{4log3}) of the elementary step $\Delta a$ in the area spectrum that we have derived  is very special. It  provides some known
relation between quantum gravity and the theory of the quasi-normal
modes of Schwarzschild black hole \cite{Hod}. The shortest link is
the following  definition of a frequency $\omega$,
\begin{equation}
\omega\ :=\ \frac{\Delta a}{8 \sqrt{\pi a}}\label{Momega}
\end{equation}
($G=\hbar=1$ in this
subsection). The frequency $\omega$ given by $\Delta a=4\log 3$ turns
out to coincide with the limit of the quasi-normal mode frequencies
at which the dumping is maximal. Moreover, equation (\ref{Momega}) has a
heuristic quantum gravitational derivation \cite{Hod}:  the quantum
jump $a\mapsto a+\Delta a$ is assumed to be caused by the quantum
excitation $M\mapsto M+\omega$ of the black hole mass $M$, where
$$a\ =\ 16\pi M^2$$
implies
$$\Delta a \ =\ 32\pi M\omega.$$

It was also noticed \cite{Dreyer} that the value $4\log 3$ for
$\Delta a$ could be predicted by the ABCK model with the standard
LQG quantum area operator (\ref{LQGarea}), provided the following
two assumptions were satisfied
\begin{itemize}
\item the quantum numbers $j_i$ in (\ref{bulkstates}) take integer values
only (rather then half integers),
\item the states corresponding to the lowest non-zero value of $j_i$
contribute  to the entropy in the leading order as
$a\rightarrow\infty$.
\end{itemize}
The first assumption would be equivalent  to replacing $SU(2)$ by
$SO(3)$ in LQG. The second assumption had been believed in for a while,
but turned out to be incorrect \cite{DL}. With the quantum flux-area operator, on the
other hand, the result (\ref{4log3})  holds with the $SU(2)$ group and
is derived by an exact calculation.

Remarkably, that result (\ref{4log3}) is in fact insensitive to
replacing $SU(2)$ by $SO(3)$. Indeed, the replacement  amounts to
restricting the combinatorial black hole entropy   Definition
\ref{def_entropy} to the set of integer valued sequences
$(m_1,...,m_n)$. All the sequences $(m^*_1,\ldots,m^*_n)$ of this $SO(3)$ theory can be obtained out of the sequences $(m_1,\ldots,m_n)$ of our $SU(2)$ theory
by the 1 to 1 transformation
\begin{align*}
(m_1,\ldots,m_n)\ &\mapsto\ (m^*_1,\ldots,m^*_n)\ =\ (2m_1,...,2m_n), \\
\gamma\ &\mapsto\ \gamma^*\ =\ \frac{1}{2}\gamma, \\
\kappa\ &\mapsto\ \kappa^*\ =\ 2\kappa.\label{tran}
\end{align*}
Indeed, the transformation preserves the area
$$    4\pi\gamma \kappa =\ a\ = 4\pi\gamma^*\kappa^* $$
and maps the conditions
$$ 2\sum_{i=1}^n m_i\ =\ 0 \ \ ({\rm mod}\, \kappa)\,, \ \ \ \sum_{i=1}^n |m_i|\ =\ \frac{\kappa}{2}$$
into
$$ 2\sum_{i=1}^n m^*_i\ =\ 0 \ \ ({\rm mod}\, \kappa^*)\,, \ \ \ \sum_{i=1}^n |m^*_i|\ =\ \frac{\kappa^*}{2}.$$
As a consequence, the transformation preserves the entropy function $a\mapsto S_{\rm bh}(a)$.
Since it also preserves the spectrum of the quantum flux-area operator and the quantity $\Delta a$, the mysterious relation with the quasi-normal modes is maintained.

\section{Comments and conclusions}{\label{comm}}

The main idea presented in the paper is to substitute the area operator used in the definition of black hole entropy according to the ABCK prescription for a different area operator $\hat{a}^{\rm flux}$ that can be defined in the LQG framework by using some extra structure provided by the inner spacetime boundary introduced to model a black hole. Other than this we strictly adhere to the entropy definition of \cite{ABK}. The use of this new operator has several advantages. First of all its spectrum is equally spaced and, furthermore, the prequantized values of the area belong to the spectrum. This reinforces the beautiful interplay between the bulk quantum geometry and the horizon CS theory shown in \cite{ABCK,ABK} and allows us to eliminate the need to introduce an area interval of undetermined width $\delta$ at this stage of the definition of the entropy. As we have shown, it is possible to completely solve the problem of determining the black hole entropy with the new area operator. Actually we have been able to get the solution in a closed (and rather simple) form and also obtain the asymptotic behavior for large areas.

In order to compare our intrinsically quantum entropy with the semiclassical Bekenstein-Hawking entropy, we course grain the quantum entropy along an interval of arbitrary width $\Delta$ greater then then the area gap. An agreement is established if and only if the Immirzi parameter is $\gamma=(\log 3)/\pi$.

The main drawback of our choice is that the  law is not immediately recovered due to the different behavior of the entropy for ``even'' and ``odd'' areas. Within each of these subsets of area eigenvalues the right proportionality between entropy and area can be found after fixing an appropriate value for  $\gamma_{\rm even}$ or $\gamma_{\rm odd}$ so if one can physically justify the elimination of one of these two sectors in the spectrum of the area operator one would get the desired result without any coarse graining. Given that the spacing between consecutive eigenvalues depends on the choice of $\gamma$ it is actually possible to have the type of area quantization proposed in \cite{BM} with a distance between consecutive eigenvalues of an integer multiple of $4\log 2$ or connect with some proposals involving the choice of $SO(3)$ as the internal symmetry group \cite{soo,Dreyer}. Another potential problem that should be faced at some point is the recovery of a thermal spectrum for the Hawking radiation with such an equally spaced spectrum \cite{BM2}.

Proposals with a flavor similar to ours have appeared in the literature before. For example Krasnov has proposed in \cite{Krasnov} to use an operator of the type
\begin{eqnarray*} \hat{a}^{\rm Kr}\,|(m_1,j_1,\ldots,m_n,j_n),\cdots\rangle_{\rm Bul} &=&
8\pi\gamma \ell^2_{P}\sum_{i=1}^n j_i\,
|(m_1,j_1,\ldots,m_n,j_n),\cdots\rangle_{\rm
Bul}
\end{eqnarray*}
that can be understood as an approximation to the standard area operator in LQG justified by the fact that for large $j$ we have $\sqrt{j(j+1)}\sim j$. It is important to notice that although the spectrum of this area operator is the same as that of $\hat{a}^{\rm flux}$ its introduction in the ABCK entropy definition would require to keep the necessity to define compatible lists of $j$'s that are not needed in our case. It is rather straightforward to adapt our methods to this case. When this is done the result for odd areas is the same as the one that we give here whereas for even areas is asymptotically the same.

Another such study is due to Sahlmann \cite{Hanno}. In that paper the author starts from the Domagala-Lewandowski prescription \cite{DL} to obtain the entropy by studying only sequences of third spin components. His starting point is to count $m$-sequences satisfying an area constraint with the form of an inequality
$$
\sum_{i=1}^n\sqrt{|m_i|(|m_i|+1)}\leq \frac{a}{2}
$$
and the additional projection constraint given by
$$
\sum_{i=1}^n m_i=0.
$$
He then proceeds to approximate each $\sqrt{|m_i|(|m_i|+1)}$ term as $|m_i|$ or $|m_i|+1/2$ and solve the resulting problem by using generating function techniques similar to the ones that we have used here. In fact, as mentioned in appendix \ref{AppGen}, some of the results that we have used to compute the entropy actually appear in \cite{Hanno}.

To conclude, in our opinion the framework that we have developed here provides a concrete and carefully defined starting point to attack the problem of understanding Hawking radiation in LQG. We hope that the precise knowledge of the degeneracies in the spectrum of the area operator and the mathematical tools used here will allow us to solve this problem and find out if the description presented here is really suitable to model black holes.

\bigskip

\noindent{\bf Acknowledgements} We would like to thank John Baez for
pointing out the relation  with the quasi normal modes. The work was
partially supported by the Spanish MICINN research grant FIS2008-03221, the Consolider-Ingenio 2010 Program CPAN (CSD2007-00042),
the Polish Ministerstwo Nauki i Szkolnictwa
Wyzszego grants 1 P03B 075 29, 182/N-QGG/2008/0, and by 2007-2010
research project N202 n081 32/1844, the National Science Foundation
(NSF) grant PHY-0456913 and by the Foundation for Polish Science
grant "Master". F. Barbero acknowledges the financial support provided by the European Science Foundation Short Visit Grant `A new approach to the study of black hole entropy in loop quantum gravity' within the framework of the activity titled `Quantum Geometry and Quantum Gravity'.

\appendix

\section{Partitions of an integer}{\label{AppPartitions}}

In this appendix we give some results used in the body of the paper, concerning the partitions of a positive integer $\kappa\in \mathbb{N}$. The following theorem is a well-known result in Combinatorics.
\begin{thr}[partitions of an integer]{\label{partitions}}
The number of (unordered) partitions of $\kappa\in\mathbb{N}$ is
$$
[x^\kappa]\prod_{i=1}^\infty \frac{1}{1-x^i}\,.
$$
The number of ordered partitions of $\kappa$ is
$$
[x^\kappa]\frac{1}{1-\sum_{i=1}^\infty x^i}=[x^\kappa]\frac{1-x}{1-2x}=2^{\kappa-1}\,.
$$
\end{thr}
Here $[x^\kappa]f(x)$ denotes the coefficient of the $x^\kappa$ term in the Taylor series expansion of the function $f$ around $x=0$.

\bigskip

The sets $\mathcal{V}_\kappa^+$ and $\mathcal{V}_\kappa^-$, defined by
 $$
\mathcal{V}^{\pm}_\kappa=\Big\{\nu\,:\, \exists n>2\,,\, \nu=(\nu_1,\ldots,\nu_n)\in \mathbb{Z}_*^n\,,\, \sum_{i=1}^n|\nu_i|= \kappa, \sum_{i=1}^n \nu_i=\pm \kappa\Big\}\,,
 $$
 are obviously bijective with the set of those ordered partitions of $\kappa$ different from the trivial one provided by $\kappa$ itself.  Hence, as a corollary of theorem \ref{partitions}, we get
$$
|\mathcal{V}_\kappa^+|=|\mathcal{V}_\kappa^-|=2^{\kappa-1}-1\,.
$$
Theorem \ref{partitions} can be easily generalized to account for the class of ordered partitions related to the black hole entropy computations. In particular:
\begin{thr}{\label{partitions2}}
The number of ordered partitions of an even integer $\kappa=2u$ for which all parts are different from $u$ is given by
$$
[x^{2u}]\frac{1}{1-\sum_{i=1}^\infty x^i+x^u}=[x^{2u}]\frac{1-x}{1-2x+x^u-x^{u+1}}=2^{2u-1}-2^{u-2}(u+3)+1\,.
$$
\end{thr}
In section \ref{counting} we introduced the sets
$$
\tilde{\mathcal{V}}^\pm_{2u}=\{\nu\in\mathcal{V}^\pm_{2u}\,:\, \nu_i\neq u,\, \forall i \}\,.
$$
Clearly,  $\tilde{\mathcal{V}}^\pm_{2u}$ are bijective with the set of those ordered partitions of $\kappa=2u$ that are different from the trivial one $(2u)$ and for which all parts are different form $u$. Hence, as a corollary of theorem \ref{partitions2}, we have
$$
|\tilde{\mathcal{V}}^\pm_{2u}|=2^{2u-1}-2^{u-2}(u+3)\,.
$$

\section{Generating function for $|\mathcal{V}^0_\kappa|$}{\label{AppGen}}
The cardinality of the set
$$
\mathcal{V}^0_\kappa:=\{\nu\,:\, \exists n\in \mathbb{N}\,,\, \nu=(\nu_1,\ldots,\nu_n)\in \mathbb{Z}_*^n\,,\quad \sum_{i=1}^n|\nu_i|= \kappa, \sum_{i=1}^n \nu_i=0\}
$$
can be computed by considering the more general sets
$$
\mathcal{V}^p_\kappa:=\{\nu\,:\, \exists n\in \mathbb{N}\,,\, \nu=(\nu_1,\ldots,\nu_n)\in \mathbb{Z}_*^n\,,\quad \sum_{i=1}^n|\nu_i|= \kappa, \sum_{i=1}^n \nu_i=p\}
$$
corresponding to \textit{projection constraints}  $\sum_{i=1}^n \nu_i=p$, $p\in \mathbb{Z}$. Notice that the sets $\mathcal{V}^\pm_\kappa$ defined in (\ref{nupm}) are related, but not equal due to the trivial sequences $(\pm\kappa)$, to  $\mathcal{V}^{\pm\kappa}_\kappa$. The generating function of the double sequence  $\{|\mathcal{V}^p_\kappa|\,:\, k\in \mathbb{N},p\in \mathbb{Z}\}$ can be obtained by particularizing the results of \cite{EF} to the easier case corresponding to the spectrum of the flux-area operator (\ref{fluxarea}).

\bigskip

\begin{thr}[generating function for the numbers $|\mathcal{V}^p_\kappa|$]{\label{Nkp}} The function
\begin{eqnarray*}
G(x,z):=\frac{z-x-z^2x+zx^2}{z-2x-2z^2x+3zx^2}
\end{eqnarray*}
satisfies
\begin{eqnarray*}
G(x,z)=1+\sum_{\kappa\in \mathbb{N}}\Big(\sum_{p\in \mathbb{Z}}|\mathcal{V}^p_\kappa|\,z^p\,\Big)\,x^\kappa \,.
\end{eqnarray*}
\end{thr}
\begin{proof}
 The proof is a trivial generalization of the methods developed in \cite{EF}.
The formulas appearing in \cite{EF} are related to the problem of counting the number of solutions to certain quadratic diophantine equations. Here, the role of those Pell equations is played by the ones giving the number of partitions for certain integers. This gives rise to the terms $\sum_{\kappa=1}^\infty x^\kappa$ in the generating functions. On the other hand, the projection constraint $\sum_i \nu_i=p$ can be incorporated by using a generating function in the form of a Laurent  polynomial in an auxiliary variable $z$. Actually,
\begin{eqnarray*}
G(x,z)&=&\left(1-\sum_{n=1}^\infty(z^n+z^{-n}) x^n\right)^{-1}=\frac{z-x-z^2x+zx^2}{z-2x-2z^2x+3zx^2}
\end{eqnarray*}
satisfies $
|\mathcal{V}^p_\kappa|=[z^p] [x^\kappa]G(x,z)$ whenever $\kappa\in \mathbb{N}$ and $p\in \mathbb{Z}$. The notation $[z^p] [x^\kappa]G(x,z)$ means that we must perform first the  power expansion of $G(x,z)$ in $x$ and then find the Laurent expansion in $z$ of the coefficients previously obtained. The normalization $G(0,z)=1$ is chosen to account for the trivial sequence appearing in definition \ref{def_entropy}.
\end{proof}

\bigskip

The problem of finding generating functions for the combinatorial problems posed above was considered  also by H. Sahlmann in \cite{Hanno}. His approach was based on the use of paths in the integer lattice $\mathbb{Z}$. The generating function $G_{\mathrm{HS}}(x,z)$  provided in page 11 of reference \cite{Hanno} --we are using  $G_{\mathrm{HS}}(x,z)$ for Sahlmann's  $G(g,z)$-- can be obtained from $G(x,z)$ as
$$
G_{\mathrm{HS}}(x,z)=G(x,z)-1=\frac{x(1+z^2-2xz)}{z-2x-2z^2x+3zx^2}\,.
$$
This is so because $G_{\mathrm{HS}}$ satisfies $G_{\mathrm{HS}}(0,z)=0$. As we can see the different approaches provide equivalent results.

\bigskip

By using Cauchy's theorem, it is possible to integrate out the auxiliary $z$-variable appearing in $G(x,z)$ and obtain a generating function $G^0(x)$ for the numbers $|\mathcal{V}^0_\kappa|$.

\begin{thr}[generating function for $|\mathcal{V}^0_\kappa|$]{\label{Nk0}}
The function \begin{eqnarray}
G^0(x)=\frac{1}{2}\left(1+\sqrt{\frac{1-x^2}{1-9x^2}}\right)\label{G0}
\end{eqnarray}
satisfies
$$
G^0(x)=1+\sum_{\kappa\in\mathbb{N}}|\mathcal{V}^0_\kappa|x^\kappa\,.
$$
Hence
\begin{eqnarray*}
|\mathcal{V}^0_{2u}|=\frac{(-1)^u \sqrt{\pi}}{4 \Gamma(3/2 - u)\cdot u!}\, _{2}F_{1}(1/2, -u; 3/2 - u; 9)\,,\quad
|\mathcal{V}^0_{2u+1}|&=&0\,,
\end{eqnarray*}
where $\Gamma$ and $_{2}F_{1}$ are, respectively, the Gamma function and the Gauss's hypergeometric function.
\end{thr}

\begin{proof} Let
\begin{eqnarray*}
x_{\pm}(z)&:=&\frac{1+z^2\pm\sqrt{1-z^2+z^4}}{3z}\,,\quad
z_{\pm}(x):=\frac{1+3x^2\pm\sqrt{1-10x^2+9x^4}}{4x}\,.
\end{eqnarray*}
By using theorem \ref{Nkp}, it is possible to write
\begin{eqnarray*}
|\mathcal{V}^0_\kappa|&=&[z^0][x^\kappa]G(x,z)\\
&=&\frac{1}{2\pi i}\oint_{C_1}\left(\frac{1}{2\pi i} \oint_{C_0} \frac{z-x-z^2x+zx^2}{(1-x)(z-2x-2z^2x+3zx^2)}\frac{\mathrm{d}x}{x^{\kappa+1}}\right)\, \frac{\mathrm{d}z}{z}\,, \quad \kappa\in \mathbb{N}\,,
\end{eqnarray*}
where $C_0$ and $C_1$ are simple closed curves in the complex plane satisfying the following conditions. Both curves must surround the origin. The curve $C_0$ has to enclose $x_0=0$ but not  $x_{\pm}(z)$, for all $z\in C_1$. It is not difficult to prove that this last requirement implies that $C_1$ must enclose both $z_0=0$ and $z_-(x)$ but not $z_+(x)$, for all $x\in C_0$. If we use now Fubini's theorem
$$
|\mathcal{V}^0_\kappa|=\frac{1}{2\pi i}\oint_{C_0}\left(\frac{1}{2\pi i} \oint_{C_1} \frac{z-x-z^2x+zx^2}{(1-x)(z-2x-2z^2x+3zx^2)}\frac{\mathrm{d}z}{z}\right)\, \frac{\mathrm{d}x}{x^{\kappa+1}}\,.
$$
Hence
\begin{eqnarray*}
G^0(x)&:=&\frac{1}{2\pi i}\oint_{C_1} \frac{ G(x,z)}{z}\,\mathrm{d}z=\frac{1}{2\pi i}\oint_{C_1}
\frac{z-x-z^2x+zx^2}{z(z-2x-2z^2x+3zx^2)}
\,\mathrm{d}z
\end{eqnarray*}
satisfies
$$
|\mathcal{V}^0_\kappa|=[x^\kappa]G^0(x)\,,\quad \kappa\in \mathbb{N}\,.
$$
The function $G(x,z)/z$ has simple poles at $z_0$ and $z_\pm(x)$ and their corresponding residues are
$$
\mathrm{Res}(G(x,z)/z,z_0)=\frac{1}{2}\,,\quad \mathrm{Res}(G(x,z)/z,z_\pm(x))=\mp\frac{1-x^2}{2\sqrt{1-10x^2+9x^4}}\,.\quad
$$
Therefore, applying Cauchy's theorem,
\begin{eqnarray*}
G^0(x)&=&\frac{1}{2}+\frac{1-x^2}{2\sqrt{1-10x^2+9x^4}}=\frac{1}{2}\left(1+\sqrt{\frac{1-x^2}{1-9x^2}}\right)\,.
\end{eqnarray*}
Finally, by using the binomial formula and the standard properties of the $_{2}F_{1}$ hypergeometric functions, we find
\begin{eqnarray*}
G^0(x)&=&\frac{1}{2}\left(1+\sqrt{\frac{1-x^2}{1-9x^2}}\right)
\\&=&1+\frac{1}{2}\sum_{u=1}^\infty (-1)^u\left(\,\sum_{m=0}^u 3^{2m}\binom{1/2}{u-m}\binom{-1/2}{m}\right)\, x^{2u}\\&=&
1+\sum_{u=1}^\infty \frac{(-1)^u \sqrt{\pi}}{4 \Gamma(3/2 - u)\cdot u!} \, _{2}F_{1}(1/2, -u; 3/2 - u; 9)\, x^{2u}\,.
\end{eqnarray*}
\end{proof}

\section{Asymptotic expansion of $_{2}F_{1}$}{\label{AppAsym}}

In this appendix we want to briefly discuss how the asymptotic expansion of the entropy is obtained. To this end we need to find out the asymptotic behavior of a function $f$ defined in terms of a hypergeometric function $_{2}F_{1}$ in  the form $f(u)=\,_{2}F_{1}\left(\alpha,\beta-u;\gamma-u;x\right)$ with $\alpha$, $\beta$, $\gamma$, and $x$ fixed and $u\rightarrow\infty$. These expansions were discussed in detail in a classic paper by G. N. Watson \cite{Watson}. In order to apply the techniques developed in that article it is necessary to make use of some identities relating hypergeometric functions of different arguments. This is done in another classic book by A. R. Forsyth \cite{Forsyth} from which the notation used by Watson in \cite{Watson} is borrowed. Specifically,
because of the following identity
\begin{eqnarray*}
& &\frac{(-1)^u \sqrt{\pi}}{4\Gamma(3/2-u) \Gamma(u+1)}\,_{2}F_{1}\left(\frac{1}{2},-u;\frac{3}{2}-u;9\right)=
\frac{3^{2u-1}\Gamma(u)\sqrt{2}}{\Gamma(u+1/2)\sqrt{\pi}}
\,_{2}F_{1}\left(-\frac{1}{2},\frac{1}{2};\frac{1}{2}+u;\frac{9}{8}\right)\\
&&\hspace{3cm}+\frac{(-1)^{u+1}i\Gamma(u)\sqrt{\pi}}{8\Gamma(u+1/2)\Gamma(1/2-u)\Gamma(u+3/2)\sqrt{2}} \,_{2}F_{1}\left(\frac{3}{2},\frac{1}{2};\frac{3}{2}+u;-\frac{1}{8}\right)\,,
\end{eqnarray*}
the relevant asymptotic expansion can be obtained directly from the asymptotics of hypegeometric funtions when only one of the parameters goes to infinity. In our case we just have to use the formula given in \cite{Watson}
$$
_{2}F_{1}\left(\alpha,\beta;\gamma+u;x\right)\sim \frac{\Gamma(\gamma+u)}{\Gamma(\gamma+u-\beta)u^\beta}\bigg(1+O(1/u)\bigg),\quad u\rightarrow\infty\,.
$$
This way we finally get
\begin{eqnarray}
\frac{(-1)^u \sqrt{\pi}}{4\Gamma(3/2-u) \Gamma(u+1)}\,_{2}F_{1}\left(\frac{1}{2},-u;\frac{3}{2}-u;9\right)\sim 3^{2u-1}\sqrt{\frac{2}{\pi u}}\Big(1+O(1/u)\Big)\,.
\end{eqnarray}

\section{Examples}{\label{examples}}

We give here two explicit examples, corresponding to odd and even values of $\kappa$ to illustrate the main results and derivations in the paper, in particular the problems associated to the overcounting occurring when $\nu$-sequences are used instead of $b$-sequences.

\subsection{An odd $\kappa$: the case $\kappa=5$}

In this case
\begin{eqnarray*}
\mathcal{V}_5&=&\Big\{\nu\,:\, \exists n\in \mathbb{N}\,,\, \nu=(\nu_1,\ldots,\nu_n)\in \mathbb{Z}_*^n\,,\, |\nu_i|\neq 5\,,\, \sum_{i=1}^n|\nu_i|= 5,\,  \sum_{i=1}^n \nu_i=0\textrm{ or }\pm 5\Big\}\,.
\end{eqnarray*}
We can enumerate the elements of this set by first finding all the partitions of the number 5. These are: $5=4+1=3+2=3+1+1=2+2+1=2+1+1+1=1+1+1+1+1$. The simple parity argument given in the text shows that the set $\mathcal{V}_5^0$ is empty so we have
$$
\mathcal{V}_5=\mathcal{V}_5^+\cup\mathcal{V}_5^-.
$$
with the disjoint sets $\mathcal{V}_5^+$ and $\mathcal{V}_5^-$ given by
\begin{eqnarray*}
\mathcal{V}_5^+&\!\!=\!\!&\{
\mathbf{(4,1)},\mathbf{(1,4)},\mathbf{(3,2)},\mathbf{(2,3)},\\
& & (3,1,1),(1,3,1),(1,1,3),(2,2,1),(2,1,2),(1,2,2),\\
&&(2,1,1,1),(1,2,1,1),(1,1,2,1),(1,1,1,2),(1,1,1,1,1)
\}\\
\mathcal{V}_5^-&\!\!=\!\!&\{
\mathbf{(-1,-4)},\mathbf{(-4,-1)},\mathbf{(-2,-3)},\mathbf{(-3,-2)},\\
& & (-3,-1,-1),(-1,-3,-1),(-1,-1,-3),(-2,-2,-1),(-2,-1,-2),(-1,-2,-2),\\
&&(\!-2,\!-1,\!-1,\!-1),(\!-1,\!-2,\!-1,\!-1),(\!-1,\!-1,\!-2,\!-1),(\!-1,\!-1,\!-1,\!-2),(\!-1,\!-1,\!-1,\!-1,\!-1)
\}.
\end{eqnarray*}
We explicitly see here several features of these sets already commented in the text; for example there are no unit-length sequences, all the elements in  $\mathcal{V}_5^+$ are strictly positive and strictly smaller than 5 and there are no equivalent sequences within each of them (similarly for $\mathcal{V}_5^-$). The first four elements of each set (highlighted by using a boldface type) are pairwise equivalent: i.e. $(4,1)\equiv(-1,-4)$, $(1,4)\equiv(-4,-1)$ and so on. These type of sequences are characterized in proposition \ref{propV3}. As we can see there are $2u=4$ such pairs that correspond only to 4 $b$-sequences. Taking this into account we conclude that
\begin{eqnarray*}
\mathcal{B}_5&\!\!=\!\!&\{\mathbf{([4]_5,[1]_5),([1]_5,[4]_5),([3]_5,[2]_5),([2]_5,[3]_5),}\\
& & ([3]_5,[1]_5,[1]_5),([1]_5,[3]_5,[1]_5),([1]_5,[1]_5,[3]_5),([2]_5,[2]_5,[1]_5),([2]_5,[1]_5,[2]_5),
([1]_5,[2]_5,[2]_5),\\
& & ([2]_5,[4]_5,[4]_5),([4]_5,[2]_5,[4]_5),([4]_5,[4]_5,[2]_5),([3]_5,[3]_5,[4]_5),([3]_5,[4]_5,[3]_5),
([4]_5,[3]_5,[3]_5),\\
&&([2]_5,[1]_5,[1]_5,[1]_5),([1]_5,[2]_5,[1]_5,[1]_5),([1]_5,[1]_5,[2]_5,[1]_5),([1]_5,[1]_5,[1]_5,[2]_5),
\\
&&([3]_5,[4]_5,[4]_5,[4]_5),([4]_5,[3]_5,[4]_5,[4]_5),([4]_5,[4]_5,[3]_5,[4]_5),([4]_5,[4]_5,[4]_5,[3]_5),\\
& &
([1]_5,[1]_5,[1]_5,[1]_5,[1]_5),([4]_5,[4]_5,[4]_5,[4]_5,[4]_5)\},
\end{eqnarray*}
where we have denoted $\mathbb{Z}_5=\{[0]_5,[1]_5,[2]_5,[3]_5,[4]_5\}$. Finally the entropy is
$$
S(a_5)=\log(|\mathcal{B}_5|+1)=\log(27)=\log(2^5-5),
$$
which is the result given by the formulas derived in the text.

\subsection{An even $\kappa$: the case $\kappa=4$}

In this case $u=2$ and
\begin{eqnarray*}
\mathcal{V}_4&=&\Big\{\nu\,:\, \exists n\in \mathbb{N}\,,\, \nu=(\nu_1,\ldots,\nu_n)\in \mathbb{Z}_*^n\,,\, |\nu_i|\neq 4\,,\, \sum_{i=1}^n|\nu_i|= 4,\,  \sum_{i=1}^n \nu_i=0\textrm{ or }\pm 4\Big\}\,.
\end{eqnarray*}
We can enumerate the elements of this set by first finding all the partitions of the number 4. These are now: $4=3+1=2+2=2+1+1=1+1+1+1$. We have then
$$
\mathcal{V}_4=\mathcal{V}_4^0\cup\mathcal{V}_4^+\cup\mathcal{V}_4^-,
$$
where $\mathcal{V}_4^0$ is non-empty in this case. The disjoint sets $\mathcal{V}_4^0$, $\mathcal{V}_4^+$ and $\mathcal{V}_4^-$ are given by
\begin{eqnarray*}
\mathcal{V}_4^0&=&\{(\underline{2,-2}),(\underline{-2,2}),(-2,1,1),(2,-1,-1),(1,-2,1),(-1,2,-1),(1,1,-2),(-1,-1,2)\\
 & & (1,1,-1,-1),(1,-1,1,-1), (1,-1,-1,1), (-1,1,1,-1),(-1,1,-1,1), (-1,-1,1,1)\}\\
\mathcal{V}_4^+&=&\{(3,1),(1,3),(\underline{2,2}),(2,1,1),(1,2,1),(1,1,2),(1,1,1,1)\}\\
\mathcal{V}_4^-&=&\{(-3,-1),(-1,-3),(\underline{-2,-2}),(-2,-1,-1),(-1,-2,-1),(-1,-1,-2),(-1,-1,-1,-1)\}.
\end{eqnarray*}
We can see now several features of these sets that have been discussed in the paper. First of all we see that all the elements within each of the sets $\mathcal{V}_4^+$ and $\mathcal{V}_4^-$ are inequivalent. We see that all the elements in $\mathcal{V}_4^0$ satisfy that their entries are $|\nu_i|\leq2$ as stated in proposition \ref{cota}, also the only equivalent sequences in $\mathcal{V}_4^0$ (that we have underlined) are $(2,-2)$ and $(-2,2)$ as stated in proposition \ref{p0}. Their equivalence class also contains an element in each of $\mathcal{V}_4^+$ and $\mathcal{V}_4^-$ (that we have also underlined); they are $(2,2)$ and $(-2,-2)$ that are equivalent as stated in proposition \ref{propV3}. We can readily see that the equivalence classes of $\mathcal{V}_4/\equiv$ that do not have representatives in $\mathcal{V}_4^0$ are precisely those in either $\mathcal{V}_4^+$ or $\mathcal{V}_4^-$ with no element equal to $u=2$ [for example $(1,1,1,1)$ or $(-1,-1,-1,-1)$]. Actually the equivalence classes consisting in more than one element are
\begin{eqnarray*}
&& \hspace{-3cm}([1]_4,[3]_4)=\{(-3,-1),(1,3)\}\\
&& \hspace{-3cm}([3]_4,[1]_4)=\{(-1,-3),(3,1)\}\\
&& \hspace{-3cm}([2]_4,[2]_4)=\{(2,-2),(-2,2),(2,2),(-2,-2)\}\\
&& \hspace{-3cm}([2]_4,[1]_4,[1]_4)=\{(2,1,1),(-2,1,1)\}\\
&& \hspace{-3cm}([1]_4,[2]_4,[1]_4)=\{(1,2,1),(1,-2,1)\}\\
&& \hspace{-3cm}([1]_4,[1]_4,[2]_4)=\{(1,1,2),(1,1,-2)\}\\
&& \hspace{-3cm}([2]_4,[3]_4,[3]_4)=\{(2,-1,-1),(-2,-1,-1)\}\\
&& \hspace{-3cm}([3]_4,[2]_4,[3]_4)=\{(-1,2,-1),(-1,-2,-1)\}\\
&& \hspace{-3cm}([3]_4,[3]_4,[2]_4)=\{(-1,-1,-2),(-1,-1,2)\}\\
\end{eqnarray*}
So we finally get that
 \begin{eqnarray*}
\mathcal{B}_4&=&\{([1]_4,[3]_4),([3]_4,[1]_4),([2]_4,[2]_4),([2]_4,[1]_4,[1]_4),([1]_4,[2]_4,[1]_4),
([1]_4,[1]_4,[2]_4),\\
&&
([2]_4,[3]_4,[3]_4),([3]_4,[2]_4,[3]_4),([3]_4,[3]_4,[2]_4),\\
&&([1]_4,[1]_4,[3]_4,[3]_4), ([1]_4,[3]_4,[1]_4,[3]_4),([1]_4,[3]_4,[3]_4,[1]_4), ([3]_4,[1]_4,[1]_4,[3]_4), \\ & & ([3]_4,[1]_4,[3]_4,[1]_4),([3]_4,[3]_4,[1]_4,[1]_4),([1]_4,[1]_4,[1]_4,[1]_4),([3]_4,[3]_4,[3]_4,[3]_4)\},
\end{eqnarray*}
$|\mathcal{B}_4|=17$ and the entropy is
$$
S(a_4)=\log(|\mathcal{B}_4|+1)=\log(18),
$$
which is, again, given by the formulas derived in the text.

\end{document}